# Stellar intensity interferometry over kilometer baselines:
## Laboratory simulation of observations with the Cherenkov Telescope Array

Dainis Dravins[*], Tiphaine Lagadec[**]
Lund Observatory, Box 43, SE-22100 Lund, Sweden

**ABSTRACT**

A long-held astronomical vision is to realize diffraction-limited optical aperture synthesis over kilometer baselines. This will enable imaging of stellar surfaces and their environments, show their evolution over time, and reveal interactions of stellar winds and gas flows in binary star systems. An opportunity is now opening up with the large telescope arrays primarily erected for measuring Cherenkov light in air induced by gamma rays. With suitable software, such telescopes could be electronically connected and used also for intensity interferometry. With no optical connection between the telescopes, the error budget is set by the electronic time resolution of a few nanoseconds. Corresponding light-travel distances are on the order of one meter, making the method practically insensitive to atmospheric turbulence or optical imperfections, permitting both very long baselines and observing at short optical wavelengths. Theoretical modeling has shown how stellar surface images can be retrieved from such observations and here we report on experimental simulations. In an optical laboratory, artificial stars (single and double, round and elliptic) are observed by an array of telescopes. Using high-speed photon-counting solid-state detectors and real-time electronics, intensity fluctuations are cross correlated between up to a hundred baselines between pairs of telescopes, producing maps of the second-order spatial coherence across the interferometric Fourier-transform plane. These experiments serve to verify the concepts and to optimize the instrumentation and observing procedures for future observations with (in particular) CTA, the Cherenkov Telescope Array, aiming at order-of-magnitude improvements of the angular resolution in optical astronomy.

**Keywords:** Intensity interferometry, Hanbury Brown – Twiss, Aperture synthesis, Second-order coherence, Stellar surface imaging, Exoplanet imaging, Cherenkov telescopes, Cherenkov Telescope Array

## 1. INTRODUCTION

The highest angular resolution currently realized in optical astronomy is offered by amplitude (phase-) interferometers which combine light from telescopes separated by baselines of up to a few hundred meters. Tantalizing results show how stellar disks start to become resolved, revealing stars as a diversity of individual objects, although so far feasible only for a small number of the largest ones. Several concepts have been proposed to extend such facilities to scales of a km or more, but their realization remains challenging either on the ground or in space. Limiting parameters include the requirement of optical and atmospheric stability to a small fraction of an optical wavelength, and the need to cover many interferometric baselines, given that optical light cannot be copied with retained phase, but has to be split up by beamsplitters to achieve interference among multiple telescope pairs.

Bright stars typically subtend diameters of only a few milliarcseconds (mas), requiring interferometry over many hundreds of meters or some kilometer to enable surface imaging. Telescope complexes of such extent are now being constructed on the ground as arrays of air Cherenkov telescopes. These are erected to measure brief flashes of Cherenkov light in air induced by energetic gamma rays but their optomechanical performance is fully adequate also for intensity interferometry, a technique once pioneered by Hanbury Brown and Twiss[1]. Being essentially insensitive to atmospheric turbulence, this method permits both very long baselines and observing at short optical wavelengths; only the lack of suitably large and well distributed telescopes has caused the method not to be recently pursued in astronomy.

The largest current such project is CTA, the Cherenkov Telescope Array, foreseen to have some 100 telescopes spread over a few square km[2-3]. Among its envisioned uses is also that of an intensity interferometer, correlating intensity fluctuations between numerous telescope pairs on different baselines while applying timeshifts in software to track sources across the sky[4]. Since the signals are copied electronically (with no optical connection between telescopes), there is no loss of signal when forming additional baselines between any pairs of the numerous telescopes.

[*]dainis@astro.lu.se; www.astro.lu.se/~dainis  [**]lagadec.tiphaine@gmail.com

Theoretical and numerical simulations of such observations have been made and published for several types of stellar objects[5-8]. Here, we report on the first experimental simulations of these types of observations. A laboratory setup was prepared with artificial optical sources ('stars'), observed by an array of telescopes equipped with nanosecond-resolving photon-counting solid-state detectors. The photon streams (reaching MHz levels, as for realistic telescope operations) are fed into digital correlators in real time (with nanosecond resolution, as required for actual telescope operations), computing cross correlations between the 'random' (quantum optical) intensity fluctuations simultaneously measured in many different pairs of telescopes. The degree of mutual correlation for any given baseline provides a measure of the second-order spatial coherence of the source at the corresponding spatial frequency (and thus the square of the ordinary first-order coherence and the square of the Fourier transform of the source's brightness distribution). Numerous telescope pairs of different baseline lengths and orientations fill in the interferometric ($u,v$)-plane, and the measured data provide a two-dimensional map of the second-order spatial coherence of the source, from which its image can be extracted. The experience from such laboratory experiments will serve to verify concepts for detectors and electronics and serve as an input for specifying the instrumentation and observing procedures for future full-scale observations with (in particular) CTA, the Cherenkov Telescope Array.

## 1.1 Intensity interferometry: Origin and principles

The technique of intensity interferometry was developed already long ago for the original purpose of measuring stellar sizes, and a dedicated instrument was built at Narrabri, Australia[1]. What is observed is the second-order coherence of light (i.e., that of intensity, not of amplitude or phase), by measuring temporal correlations of arrival times between photons recorded in different telescopes. This is a two-photon process and the concept is today generally acknowledged as the first experiment in quantum optics.

The name 'intensity interferometer' is somewhat of a misnomer (nothing is interfering; an early name was 'post-detection interferometry'); rather its name was once chosen for its analogy to the ordinary phase/amplitude interferometer, which had similar applications in measuring stellar sizes. A star is measured with two separate telescopes, recording the random and very rapid intrinsic fluctuations in starlight. With the telescopes close together, the fluctuations in both telescopes are correlated in time, but lose correlation for larger separations. The variation of that cross correlation with increasing telescopic distance provides the second-order spatial coherence of starlight, corresponding to the square of the visibility observed in any classical amplitude interferometer. Thus one obtains [the square of] the Fourier transform of the brightness distribution of the source, from which its spatial properties and its two-dimensional image can be retrieved.

The quantity measured in intensity interferometry is $<I_1(t) \bullet I_2(t)> = <I_1(t)><I_2(t)>(1 + |\gamma_{12}|^2)$, where $<>$ denotes temporal averaging and $\gamma_{12}$ is the mutual coherence function of light between locations 1 and 2, the quantity commonly measured in phase/amplitude interferometers (a relation that holds for each linear polarization). Compared to randomly fluctuating intensities, the correlation between intensities $I_1$ and $I_2$ is 'enhanced' by the coherence parameter and an intensity interferometer thus measures $|\gamma_{12}|^2$ with a certain electronic time resolution. This relation holds for ordinary thermal (maximum-entropy or Gaussian) light where the light wave undergoes random phase jumps on timescales of its coherence time but not necessarily for light with different photon statistics (e.g., an ideal laser emits coherent light without any phase jumps, and thus would not generate any sensible signal in an intensity interferometer).

The great observational advantage of intensity interferometry is that it is practically insensitive to either atmospheric turbulence or to telescope optical imperfections, enabling very long baselines, as well as observing at short optical wavelengths, even through large airmasses far away from zenith. Since telescopes are connected only electronically, error budgets and required precisions relate to electronic timescales of nanoseconds, and light-travel distances of tens of centimeters or meters rather than small fractions of an optical wavelength. For more detailed discussions of the principal workings of intensity interferometry, see monographs by, e.g., Labeyrie et al.[9], Saha[10] or Shih[11].

However, there is a price to be paid for this freedom from atmospheric influences. Realistic time resolutions are much longer than typical optical coherence times for broad-band light (of perhaps $10^{-14}$ s), and any measured intensity-fluctuation signal is averaged over very many coherence times. Therefore, very good photon statistics is required for a reliable determination of the smeared-out signal, requiring large photon fluxes (thus large telescopes); already the 6.5 m flux collectors in the original intensity interferometer at Narrabri were larger than any other optical telescope at that time. Although ideas for a second-generation instrument were worked out, that was never realized, partly because of the then concurrent developments of phase/amplitude interferometry. However, following this start in astronomy, the technique has been vigorously pursued in other fields, both for studying optical light in the laboratory[12], and in analyzing interactions in high-energy particle physics[13] (where the method is often referred to as 'HBT-interferometry' (for 'Hanbury Brown–Twiss') or 'femtoscopy' (i.e., going beyond 'microscopy').

Thus, intensity interferometry has not been further pursued in astronomy since a long time ago, largely due to its demanding requirements for large (and movable) optical flux collectors, spread over long baselines, and equipped with fast detectors and high-speed electronics. However, these requirements are now being satisfied through the combination of high-speed digital signal handling with the construction of telescope complexes, erected for a different primary purpose, namely to optically record atmospheric Cherenkov light for the study of energetic gamma rays.

**1.2 The potential of air Cherenkov telescopes**

The technical specifications for air Cherenkov telescopes such as the forthcoming CTA, the Cherenkov Telescope Array[2-3], are remarkably similar to the requirements for intensity interferometry. In the original Narrabri instrument, the telescopes were moved during observation to maintain their projected baseline; however electronic time delays can now be used instead to compensate for different arrival times of a wavefront to the different telescopes, removing the need for having them mechanically mobile.

The CTA will provide an unprecedented telescopic light-collecting area of some 10,000 m$^2$ distributed over a few square km. Of course, it will mainly be devoted to its main task of observing Cherenkov light in air; however several other applications have been envisioned, preferably to be carried out during nights with bright moonlight which – due to the faintness of the Cherenkov light flashes – might preclude their efficient observation. Besides intensity interferometry[4], suggested additional uses include searches for rapid astrophysical events[14-16], observing stellar occultations by distant Kuiper-belt objects[17] or as a terrestrial ground station for optical communication with distant spacecraft[18]. For practical operations, one proposed concept is to place auxiliary detectors on the outside of the Cherenkov camera cover lid, an approach already realized on some Cherenkov telescopes[14-15], and one which should not interfere with regular Cherenkov camera operations.

This potential of using Cherenkov telescope arrays for optical intensity interferometry has indeed been noticed by several[19-20]. If, e.g., baseline of 2 km could be utilized with CTA at λ 350 nm, resolutions would approach 30 μas, an unprecedented spatial resolution in optical astronomy[4-5].

## 2. SIMULATING INTENSITY INTERFEROMETRY

**2.1 Steps towards intensity interferometry with Cherenkov telescopes**

Several studies have considered the fundamental observational challenges, i.e., the overall sensitivity, limiting stellar magnitudes, and similar, and those 'global' issues now appear to be consistently and generally understood[4-5,19]. For example, current types of electronics, with current Cherenkov telescope designs should, during one night of observing, permit measurements of hotter stars down to visual magnitude about $m_V$=8, giving access to many thousands of sources.

Comprehensive Monte Carlo simulations were made by Rou et al.[8]. They numerically simulated effects of various noise sources, of background light from the night sky, effects of large collecting areas (i.e., non-negligible in comparison to the coherence pattern), and other. For example, the effect of large-size telescopes is that the signal will correspond to a convolution of the degree of coherence $|\gamma|^2$ with each of the telescope light-collection area shapes. However, in real observing, there will also arise various experimental issues, not all of which are yet well understood. Some were examined already long ago by Hanbury Brown et al., while others are specific to modern detectors and electronics (e.g., dead-time or saturation effects in detectors, digital data handling). Obviously, such issues need to be examined before reliable observations can be carried out at major facilities.

**2.2 Experimental test observations**

First full-scale experiments in connecting major Cherenkov telescopes for intensity interferometry were made by Dravins and LeBohec[21], using pairs of 12-m telescopes of the VERITAS array on Mt.Hopkins, Arizona, at baselines between 34 and 109 m. In each telescope, starlight was recorded by one photomultiplier in the center of the regular Cherenkov cameras, its signal digitized, and sent via optical cables to the control building, where they were fed into a real-time digital correlator at continuous photon-count rates up to some 30 MHz. Although no astrophysical correlations were recorded, these experiments verified that there seemed to be no particular operational problems in using large Cherenkov telescopes for intensity interferometry.

Significant efforts have been invested at The University of Utah to develop experimental test facilities for intensity interferometry, including both 3-m Cherenkov telescopes at its 'StarBase' facility and laboratory setups[22]. One unit comprises a two-channel continuous digitization system, able to stream data to computer disks for hours at 100 MHz sampling rate. In other experiments, an analog correlator based upon a Field Programmable Gate Array (FPGA) was used for laboratory tests[23] with intensity fluctuations in starlight mimicked by a pseudo-thermal light source, produced by shining a laser through a rotating ground-glass plate, producing a time-variable speckle pattern[24].

Also Pellizzari et al.[25] used a rotating ground-glass plate to create speckle patterns in the observation plane with two 14-inch (36 cm) telescopes. Cross correlating the light intensity between these telescopes provides a measure of the speckle extent, and thus of the source size. Such experiments have provided insights into detector properties and signal handling although the character of the light source only has a limited similarity to astronomical objects.

## 2.3 Photon-counting detectors

Air Cherenkov telescopes have traditionally been equipped with photomultiplier (PMT) detectors, a natural choice since the image point-spread-function in the focal plane typically is some cm in extent, well matched to PMT sizes. However, their quantum efficiency is not perfect, and their electric requirements and consequences from overexposure may cause some concerns. Recently, higher-efficiency and less fragile solid-state silicon avalanche photodiodes have been incorporated into Cherenkov telescopes[26-28] and it can be envisioned that such detectors will be used also in future telescopes, at least those with an optical two-mirror design that produce a more compact focal plane.

Single-photon-counting silicon avalanche photo-diodes (SPADs) have a potential for quantum efficiency approaching unity (and extending into the infrared), while counting individual photons at nanosecond resolution. They are operated in 'Geiger mode', where each detected photon triggers an electronic avalanche as a signature of photon detection. Following a detection, a certain deadtime occurs, during which further detections are not possible. Typical deadtimes can be on order of tens of nanoseconds, permitting count rates up to 10 MHz or somewhat higher.

In addition to dark counts, detectors also show some level of afterpulsing. There remains a possibility that an avalanche electron occasionally is caught in the potential well around some semiconductor impurity site. If that trapped electron is released after a time longer than the deadtime, it may trigger a new avalanche, correlated with the real photon event. Since, in intensity interferometry, one searches for correlated signals, such afterpulsing may appear as a false correlation. Of course, the afterpulsing is a property inside each one detector and will not (except as a small higher-order effect) be correlated between two separate ones.

Another 'peculiar' property of SPADs is the emission of light from the detector surface ('diode afterglow'). Immediately following the photon-detecting avalanche, a shower of photons is emitted *from* the detector, as it recovers. For high-speed applications in large telescopes, some such secondary light could find its way back onto the detector, causing optical 'ringing'. Such emission can also induce crosstalk between adjacent pixels in a detector array ('optical crosstalk'). In the electronic design of SPAD arrays, efforts are made to remedy such effects by, e.g., making trenches into the silicon to avoid direct exposure between adjacent pixels.

Such examples illustrate that an adequate understanding of detector properties is desirable in both the selection and use of actual detectors. Since solid-state detectors appear to be a promising choice in terms of sensitivity and handling, those were chosen for the present experiments. One practical problem is that the physical size of single-pixel detectors is tiny, only a fraction of a mm, not compatible with the large image scales in Cherenkov telescopes, where larger array detectors have to be used. Such solid-state photomultiplier arrays are now available from several manufacturers but already the experience from smaller-scale ones should be relevant for future larger-scale operations.

Single-photon-counting avalanche silicon photodiode detector modules (SPADs) from several manufacturers (*ID Quantique*, *Micro Photon Devices*, *PerkinElmer*, *SensL*) were evaluated in laboratory tests. Following those, a dozen modules from MPD, *Micro Photon Devices*, with sensitive areas of 100 μm diameter, Peltier-cooled, and with dark-count rates below 500 Hz were acquired to outfit a corresponding number of small laboratory telescopes.

## 2.4 Real-time photon correlation

An essential element of an intensity interferometer is the correlator, which provides the temporally averaged product of the intensity fluctuations $\langle \Delta I_1 \bullet \Delta I_2 \rangle$ in two telescopes, normalized by the product of average intensities in each of them, $\langle I_1 \rangle \bullet \langle I_2 \rangle$. The original interferometer at Narrabri used an analog correlator to multiply the photocurrents from its photomultipliers. Analog correlators[20,23,25] have attractive properties in often being able to handle wide electronic bandwidth. However, they may also be sensitive to electromagnetic interference, originating from all sorts of electric devices omnipresent in also observatory environments[25]. For applications in intensity interferometry, these may be critical since the measured signal is expected to be tiny, and has be segregated against a very much greater background. Of course, isolation from interference is feasible with sufficient shielding, and having optical cables instead of metallic ones may alleviate the problems, but there still remain issues at their interfaces. Because of such considerations, the signal handling was here chosen to be digital only. This may introduce limitations as to how high count rates that can be handled per channel but – since S/N is in principle independent of optical bandpass – stellar fluxes may be adjusted to manageable photon-count levels using suitably narrow-band filters.

Photon correlators are commercially available for primary applications in light scattering against laboratory specimens[12]. (Such intensity-correlation spectroscopy is the temporal analog to [spatial] intensity interferometry, and was developed after its subsequent theoretical understanding.) Based upon experiences from a series of hardware correlators from different manufacturers, current units were custom-made (by *Correlator.com*) for applications in intensity interferometry, featuring sampling frequencies up to 700 MHz, able to handle continuous photon-count rates of more than 100 MHz per channel without any deadtimes, and with on-line data transfer to a host computer. Their output contains the cross correlation function between pairs of telescopes (as well as autocorrelation functions for each of them), made up of about a thousand points. For small delays, the sampling is made with the shortest timesteps of just over 1 ns, increasing in a geometric progression to large values to reveal the full function up to long delays of even seconds or more. It is believed that their performance is adequate for full intensity interferometry experiments. In particular, it should be stressed that the correlators required here are very much more modest than those used in radio phase interferometer facilities (of supercomputer class), which not only decipher spatial but also spectral information, and with many bits of resolution.

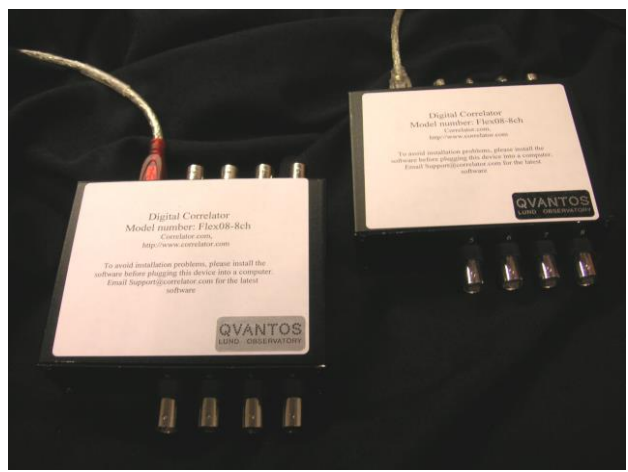

Figure 1. Digital photon correlators of the type used in present laboratory experiments. These are based upon Field Programmable Gate Arrays (FPGA) with 700 MHz clock rate, 1.4 ns time resolution, handling up to 200 MHz photon-count rates at TTL-level voltages. Each of these correlators handles signals from eight telescopes simultaneously.

Still, such correlators are not without issues. For realistic observations, the desired signal is only a small fraction of the full (Poisson-noise) intensity correlation in the raw data, and the signal must be analyzed to many decimal digits. Firmware correlators produce correlation functions in real time, processing very large amounts of photon-count data, and eliminating the need for their further handling and storage (e.g., 10 input channels, each running at 50 MHz during one 8-hour observing night process more than 10 TB of photon-count data). A disadvantage is that, if something would need to be checked afterwards, the original data are no longer available, and alternative signal processing cannot be applied.

An alternative approach is to time-tag each photon count and store all data (at least for moderate count-rates), and then later perform the correlation analyses off-line. The data streams from multiple telescopes can then be cross-correlated in software, possibly applying noise filters and also computing other spatio-temporal parameters such as higher-order correlations between three or more telescopes, which could contain additional information. Such a capability is built into the *AquEYE* and *IquEYE* instruments developed at the University of Padova for very high time-resolution astrophysics with also a view towards intensity interferometry, using similar SPAD detectors as here[29-31].

**2.5 Image reconstruction from second-order coherence**

While intensity interferometry possesses the advantage of not being sensitive to phase errors in the optical light path, ordinary two-telescope correlations also do not permit to measure such phases. These provide the absolute magnitudes of the respective Fourier transform components of the source image, while the phases are not directly obtained. Such Fourier magnitudes can well be used by themselves to fit model parameters such as stellar diameters, stellar limb darkening, binary separations, circumstellar disk thicknesses, etc., but actual images cannot be directly computed through a simple inverse Fourier transform.

While a two-component interferometer (such as the classical one at Narrabri) offers only very limited coverage of the $(u,v)$-plane, a multi-component system provides numerous baselines and an extensive coverage of the interferometric plane. Already intuitively, it is clear that the information contained there must place stringent constraints on the source image. For instance, viewing the familiar Airy diffraction pattern (cf. Figure 2 left, below), immediately recognizes it as originating in a circular aperture, although only intensities are seen. However, it is also obvious that a reasonably complete coverage of the diffraction image is required to convincingly identify a circular aperture as the source.

Various techniques (most unrelated to astronomy) have been developed for recovering the phase of a complex function when only its magnitude is known. Methods specifically for intensity interferometry were worked out by Holmes et al.[32-34] for one and two dimensions, respectively. Once a sufficient coverage of the Fourier plane is available, phase recovery and imaging indeed become possible. Nuñez et al.[6-7] applied such phase recovery to reconstruct images from simulated intensity interferometry observations, demonstrating that also rather complex images can be reconstructed (a remaining limitation is the non-uniqueness between the image and its mirrored reflection).

## 3. LABORATORY EXPERIMENTS

**3.1 Telescope array in the laboratory**

The project aim was to realize an end-to-end simulation of an intensity interferometer observing small star-like sources with a large array of telescopes, equipped with electronics of the type foreseen for future full-scale operations. An intensity interferometer was thus set up in a large optics laboratory, comprising an artificial 'star' viewed by a linear array of typically five stationary telescopes of different separations. A two-dimensional telescope layout with many tens of telescopes could be simulated by successively rotating the position angle of the 'star' relative to the plane of the telescopes. With high-speed photon-counting detectors and real-time digital cross correlation between many tens of baseline pairs, operations resemble foreseen future observations with large Cherenkov telescope arrays. Still, such a laboratory setup demands several considerations.

**3.2 Small sources and long distances**

A first issue concerns the artificial 'stars'. These should have somewhat realistic angular sizes and also possess some nontrivial properties if either single (round or elliptical), or double (with equal or unequal components). They were prepared as small apertures, drilled as physical holes in metal (using miniature drills otherwise intended for mechanical watches). This sets some minimum practical size of aperture openings on order $\sim 100$ μm. Such an opening observed at a 20 m distance subtends an angle of $5\,10^{-6}$ rad (1 arcsec). With the Airy disk diffraction radius given by $\Theta \approx 1.22\,\lambda/D$ rad, the telescope aperture D required to resolve it comes out to $\sim 10$ cm, dictating a rather compact setup of telescopes unless very large laboratory spaces are available. Our interferometer was set up in the large optics laboratory of Lund Observatory, occupying most of its wall-to-wall extent, with a source-to-telescope distance of $\sim 23$ m.

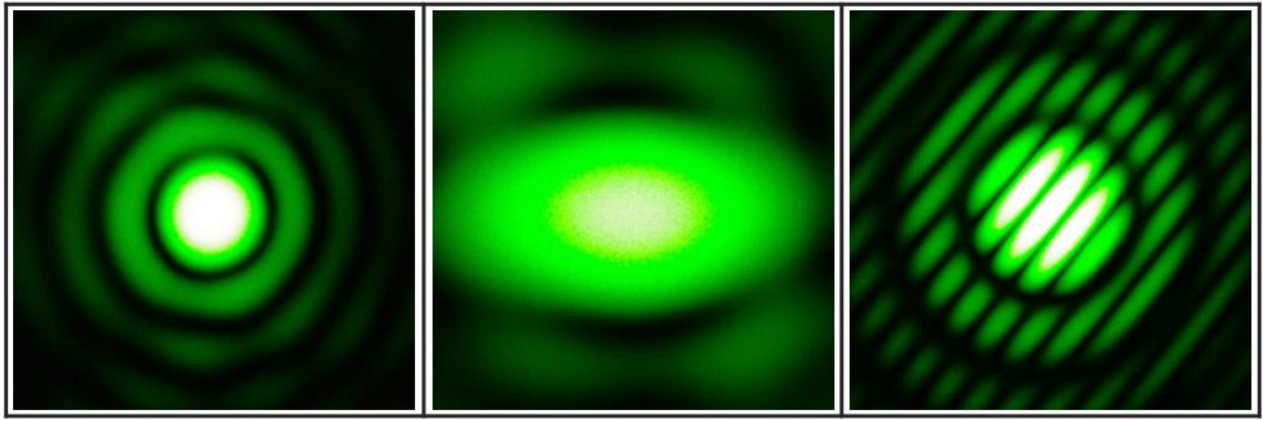

Figure 2. Diffraction patterns in λ 532 nm laser light show the [squared] Fourier transforms of some of the artificial 'stars'. Left to right: Circular single star of 150 μm nominal diameter; elliptic small single star; binary star with each component 100 μm. The side of each image corresponds to ~ 70 cm in the telescope plane and corresponding baselines are required to retrieve these patterns.

Figure 2 shows photographs of the diffraction patters of some of the artificial stars, here illuminated by coherent laser λ 532 nm light, projected onto a large screen. These patterns – Fourier transforms of the aperture openings – serve both to precisely obtain the dimensions and shapes of the 'stars' (quantities otherwise awkward to directly measure on the tiny openings) and to illustrate the spatial coherence patterns that eventually can be retrieved from interferometry.

### 3.3 Dependence on source temperature

One particular challenge, peculiar to intensity interferometry, comes from its signal-to-noise properties in measuring the second-order coherence, which for one pair of telescopes equals[1]: $(S/N)_{RMS} = A \bullet \alpha \bullet \eta \bullet |\gamma_{12}(\mathbf{r})|^2 \bullet \Delta f^{1/2} \bullet (T/2)^{1/2}$. Here $A$ is the geometric mean of the areas of the two telescopes; $\alpha$ is the quantum efficiency of the optics plus detectors; $\eta$ is the flux of the source in photons per unit optical bandwidth, per unit area, and per unit time; $|\gamma_{12}(\mathbf{r})|^2$ is the second-order coherence of the source for the baseline vector $\mathbf{r}$, with $\gamma_{12}(\mathbf{r})$ being the mutual degree of coherence; $\Delta f$ is the electronic bandwidth of the detector plus signal-handling system, and $T$ is the integration time.

Most of these parameters are readily understandable since they depend on the instrumentation, but η depends on the source itself, being a function of its brightness temperature. Thus, for a given number of photons detected per unit area and unit time, the signal-to-noise ratio is better for sources where those photons are squeezed into a narrower optical band. A corollary is one valuable property for intensity interferometry, namely that (for a flat-spectrum source) the S/N is independent of the width of the optical passband, whether measuring only the limited light inside a narrow spectral feature or a much greater broad-band flux. This property was exploited already in the original Narrabri interferometer[35] to identify the extended emission-line volume from the stellar wind around the Wolf–Rayet star $\gamma^2$ Vel. The same effect could also be exploited for increasing the signal-to-noise by observing the same source simultaneously in multiple spectral channels, a concept foreseen for the once proposed successor to the original Narrabri interferometer[36].

That the limiting S/N is set by the source temperature rather than by its brightness or telescope size, may appear somewhat counter-intuitive (and has been the cause of several misunderstandings) but can be understood both in classical and quantum terms. Once the observational equipment is in place, one may try to improve the S/N ratio. One could try increasing the photon flux by going to a broader wavelength interval, white light, say. However, the S/N will not change: realistic electronic resolutions (~ ns) are always very much slower than the temporal coherence time of broad-band light (perhaps $10^{-14}$ s). While broadening the spectral passband does increase the photon count rate, it also decreases the temporal coherence by the same factor. The intensity fluctuations have their full amplitude over one coherence time, and now get averaged over many more, canceling the effects of decreased photon noise. Then one could try with larger telescopes. However, once telescope sizes begin to approach the structures in the spatial coherence pattern, although increasing telescope areas give more photons, they also average over more spatial coherence structures by the same factor, again not improving the S/N. Alternatively, one might be tempted to observe brighter sources. However, for any given source temperature, brighter sources will be larger in angular extent and one will again average over additional spatial coherence areas, without affecting the S/N. Indeed, even pointing the telescope at the Sun will not help; the only way to get better S/N is to find sources with a higher surface brightness. Below some effective source temperature, no sensible measurements can be realized, no matter how bright the source, or how large the telescopes (Figure 3).

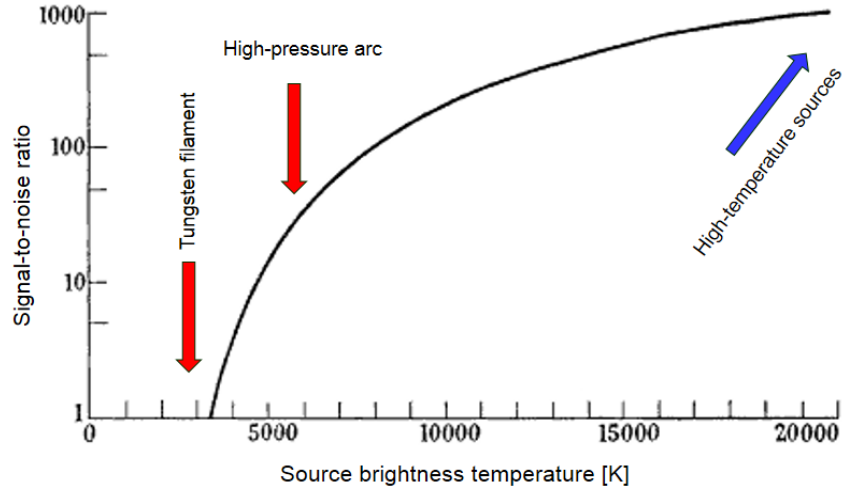

Figure 3. Temperature dependence of achievable signal-to-noise in measuring second-order coherence (assuming 100 MHz electronic bandwidth, 1-hour integration). For stars of same size but decreasing temperatures (decreasing fluxes), telescope diameter must increase to maintain the same S/N.  When the mirrors become so large that the star is resolved by a single mirror, S/N drops, and for stars cooler than a given temperature, no gain results (adapted from Hanbury Brown and Twiss[37]).

**3.4 High-temperature laboratory light sources**

That intensity interferometry is sensitive to sources of high brightness temperature but limited in observing cool ones, of course, is equally valid for any laboratory setup, as for stars in the sky. The source must be small enough to produce an extended 'diffraction' pattern that can be sampled by the interferometer while also be bright enough to produce acceptable photon count rates. However, while there are many stars in the sky with $T_{eff}$ = 10,000 K or more, to produce a correspondingly brilliant laboratory source is much more challenging. Further, the precise relation between second- and first order coherence, $g^{(2)} = 1 + |g^{(1)}|^2$, assumes that the light is chaotic (with a Gaussian amplitude distribution)[11,38-39], i.e., the light waves undergo random phase shifts so that intensity fluctuations result, which then bear a simple relation to the ordinary first-order coherence. While this must be closely satisfied for any thermal source, it is not the case for lasers which, ideally, never undergo any intensity fluctuations anywhere ($g^{(2)} = 1$). The spatial extent of a laser source therefore cannot be measured by intensity interferometry, and a laser is not an option to enhance the brightness of an artificial star.

In initial attempts to achieve a high surface brightness for an artificial 'star', the very small emission volume of a high-pressure Hg arc lamp was focused onto a pinhole serving as the 'star', and a narrow-band optical filter singled out its brightest emission line ($\lambda$ 546 nm). Although this represents about the highest blackbody brightness temperature (some 5500 K) that readily can be obtained with somewhat ordinary laboratory equipment for a non-laser source, the photon-count rates still turned out to be too low for conveniently short integration times. Other tests were made with several lamps of various atomic species, selecting their brightest and narrowest emission lines, but the signals still remained marginal. Finally, quasi-monochromatic chaotic light was produced by scattering monochromatic laser light against microscopic particles, suspended in a cuvette with room temperature water. Such particles undergo thermal (Brownian) motion, producing a slightly Doppler-broadened spectral line which is extremely narrow (~ MHz). Thus, a light source with an effective brightness temperature estimated on order $T_{eff}$ ~ 100,000 K was produced, permitting conveniently short integration times down to just tens of seconds. Such scattered laser light is used for various laboratory photon-correlation measurements of time variability[12], but we are not aware of any previous such experiment with a spatial intensity interferometer.

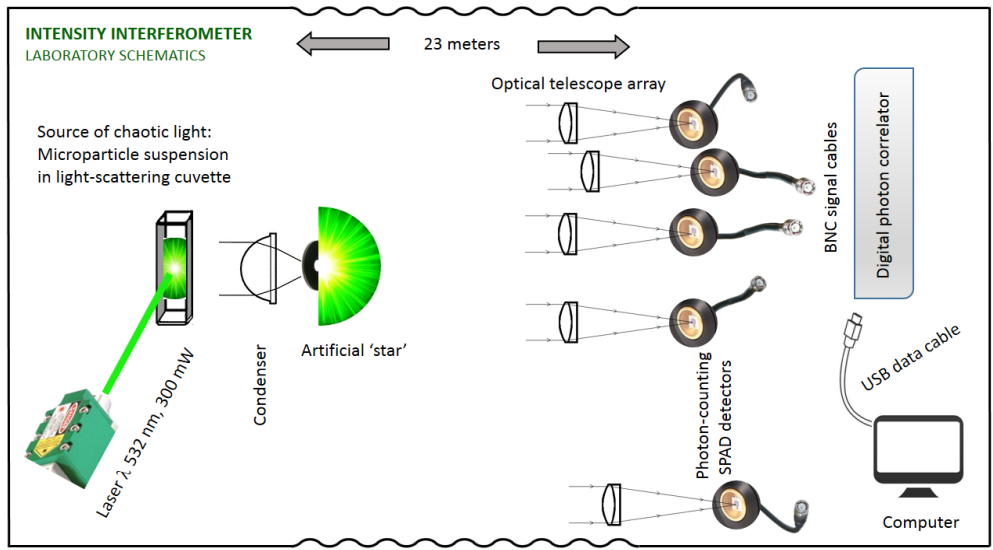

Figure 4. Intensity interferometer in the laboratory: Artificial 'stars' are illuminated by chaotic light produced by scattered laser light and observed by an array of small telescopes located at 23 m distance. The pulse streams from the photon-counting detectors are cross correlated between pairs of telescopes in a digital firmware correlator under computer control.

### 3.5 Dynamic light scattering

Monochromatic light scattered by microscopic particles in thermal (Brownian) motion is broadened by Doppler shifts, producing a spectral line with a Lorentzian wavelength shape, and with a Gaussian (chaotic, maximum-entropy) distribution of the electric field amplitudes[40-41]. Such light is thus equivalent in its photon statistics and intensity fluctuations to the thermal (white) light expected from any normal star, the difference being that the spectral passband is now very much narrower. In addition to a high $T_{eff}$, the narrowness of the spectral line implies a long coherence time, lessening the demands on the detector time resolution, and permitting measurements on also the less demanding microsecond scales, where detector issues such as afterpulsing are less pronounced. The exact values for the spectral broadening can be modified by choosing the particle size (larger particles undergo slower motion, inducing less Doppler broadening), but it also somewhat depends on the viscosity and the temperature of the medium of suspension.

Such light sources are not completely without issues, however. Intensity interferometry relies upon a relation between first- and second-order coherence that is valid for Gaussian light. The scattering against single and non-interacting particles in Brownian motion produces such light, however only in the case of single scattering. In case the photons undergo multiple scattering events, if the scattering particles experience mutual interactions, have intrinsic dynamic properties, or there are internal currents in the suspension liquid, this relation may no longer hold. In such cases, there will still be relations between $g^{(1)}$ and $g^{(2)}$, but possibly of a somewhat different form. To avoid such complications, the optical paths were minimized by focusing the laser light very close to the surface of the fluid inside the cuvette.

Chaotic light was initially produced by scattering λ 532 nm laser light against microscopic monodispersive polystyrene spheres (from *Polysciences*) suspended in a cm-sized cuvette with room temperature distilled water. These spheres are basically non-interacting particles undergoing random motion due to collisions with the water molecules that surround them. After trying out spheres of several different sizes (ranging between 50 nm and 90 µm), a diameter of 0.2 µm was chosen as giving both an acceptably high light level in the scattering, and a convenient coherence time for measurements.

However, the experiments required quite a number of tests, where the cost of volumes of such microscopic spheres became a limitation. As an inexpensive scattering fluid with comparable properties, homogenized household milk (diluted with two parts water) was then used instead. Milk contains microscopic fat globules, which undergo analogous Brownian motion, also studied with photon correlation spectroscopy[42-44]. Given that the cost per volume of milk is almost four orders of magnitude lower than that of microsphere suspensions, most experiments (including those presented here) were carried out with laser light scattering in milk. Also, milk with varying fat content was tried out, but with no obvious differences in terms of, e.g., correlation timescales. Presumably, the homogenization process produces fat globules of similar size, and the difference between milk with differing fat content lies in its quantity, not its microscopic properties.

## 3.6 Interferometer setup

Figure 4 shows the layout of the laboratory setup, and Figure 5 some photographs of its parts. The telescopes are small refractors with 25 mm diameter achromatic lenses mounted on an optical bench at some 23 meters distance from the artificial 'stars'. In each telescope, light is focused onto a SPAD, a single-photon-counting avalanche photodiode, enabling photon-count rates up to ~ 10 MHz. The pulse-train output (electronic TTL standard) is fed to a computer-controlled firmware correlator for real-time cross correlation of the data streams between various pairs of telescopes.

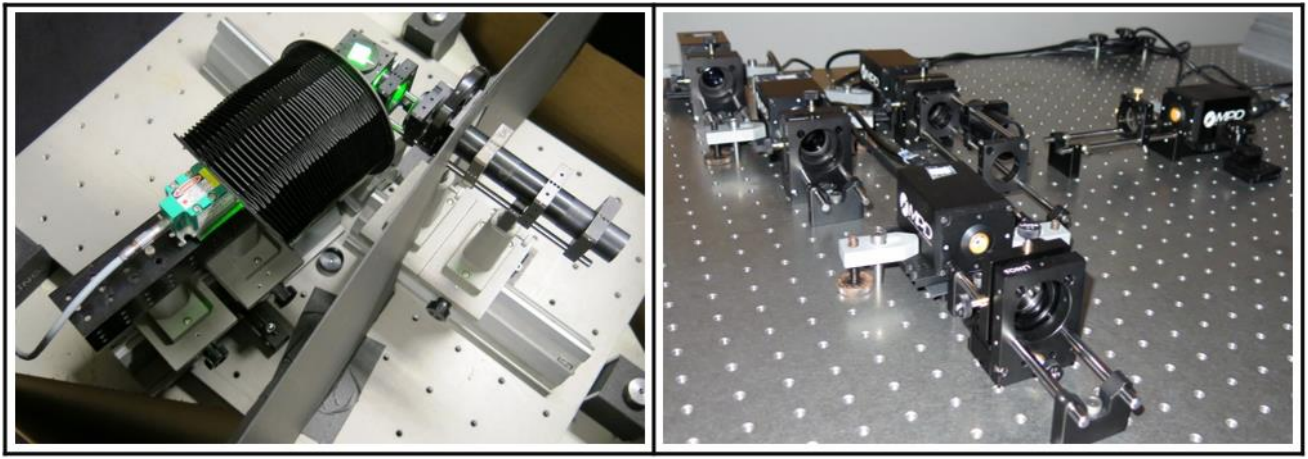

Figure 5. Components in the laboratory setup. *Left:* Light from a 300 mW λ 532 nm laser is randomized through scattering against microscopic particles in a square-top cuvette and focused by a condenser onto artificial 'stars', being apertures in a rotatable holder. *Right:* The 'stars' are observed by a group of (here) five small telescopes with 25 mm apertures, each equipped with a photon-counting SPAD detector. One unit perpendicular to the others uses a 45-degree mirror to obtain a particularly short baseline. Two-dimensional coverage is achieved by successively rotating the position angle of the source relative to the plane of the telescopes.

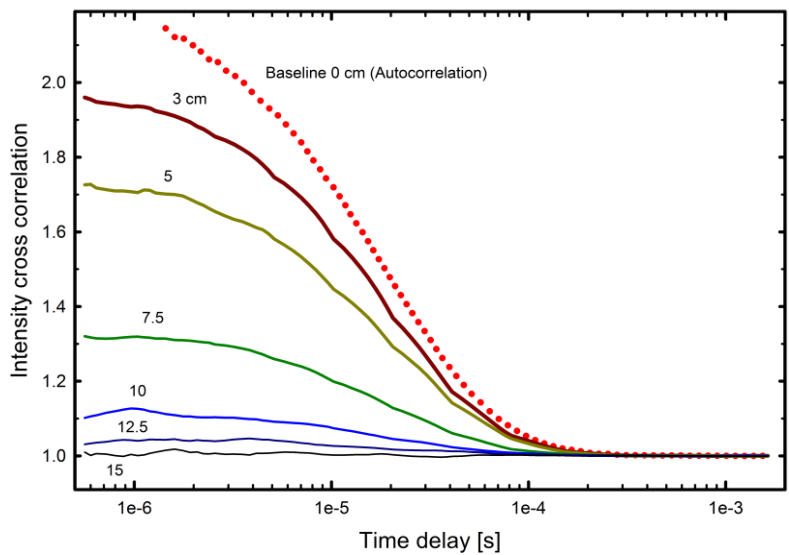

Figure 6. Cross correlation functions measured for an artificial single star of 1 arcsec apparent diameter (100 μm aperture at 23 m), normalized to zero baseline, show how (a) the temporal coherence gradually decreases with increasing time delay, and (b) the spatial coherence gradually decreases with increasing baseline. The normalized value for zero baseline is taken as the autocorrelation for delays in the range 1-10 μs (at the shortest delay times the autocorrelation rises to unphysical values due to detector afterpulsing).

## 3.7 Measurements of single 'stars'

Examples of measured intensity cross correlation functions between a pair of telescopes (1) and (2), $\langle I_1(t) \bullet I_2(t+\Delta t)\rangle$, are in Figure 6, normalized to $g^{(2)} = 2$ for zero telescope baseline (obtained as the autocorrelation in one telescope $\langle I_1(t) \bullet I_1(t+\Delta t)\rangle$). For long delay times $\Delta t$, the correlation vanishes, and the values tend to unity, as appropriate for random uncorrelated variations. The characteristic slope for delays around $\Delta t \approx 10$ μs is a measure of the temporal coherence time and indicates the optical bandpass of the scattered laser light, here on order $10^5$ Hz. For broader-band light, the coherence time is very much shorter, and then the possibilities to resolve the varying correlation for successively shorter delays become limited. However, in this arrangement it is possible to follow the changing spatial and temporal correlation in detail, as the temporal averaging proceeds over successively larger numbers of coherence times.

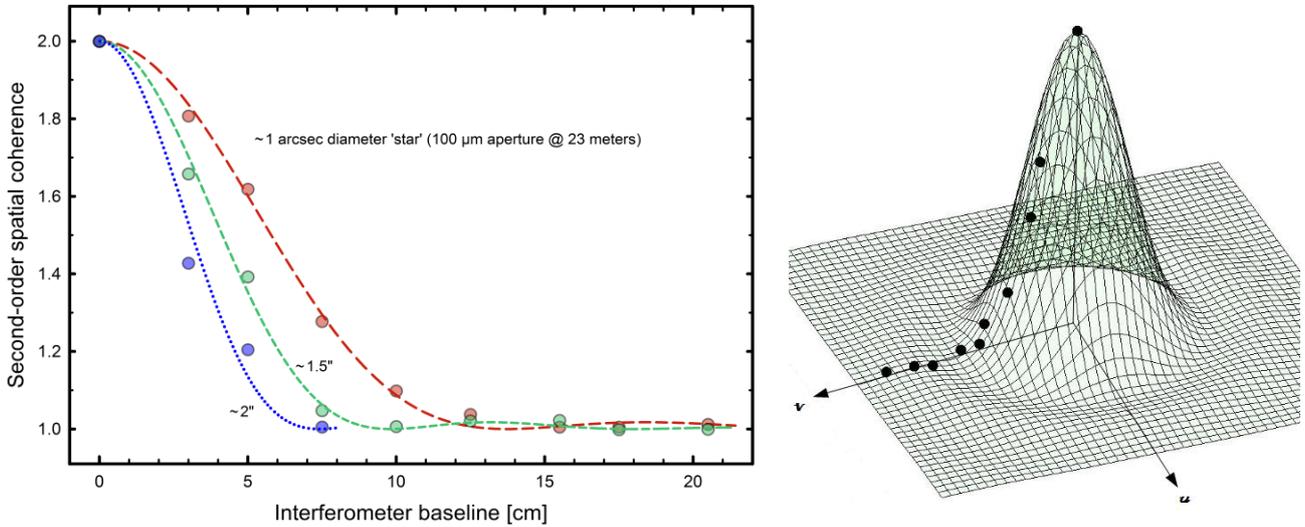

Figure 7. *Left:* Second-order coherence $g^{(2)}$ measured for artificial single stars of different angular sizes. The spatial coherence values come from correlation functions such as in Figure 6, averaged for delays between 1-10 μs, normalized to baseline zero. Superposed are theoretically expected dependences for ideal circular apertures. These Airy patterns, given by the squared moduli of the Fourier transforms, $1+ (2 J_1(x)/x)^2$, show a quite close agreement. *Right:* Such one-dimensional coherence functions sample one position angle of the two-dimensional coherence surface, here a sequence of measured points on the surface for an idealized circular aperture.

For intensity interferometry we are not concerned about the temporal coherence but the spatial one: its gradual change with increasing baseline enables the angular size of the source to be determined. At delays shorter than ~ 1 μs, the temporal coherence is here 'fully' resolved, and the differences between curves then reflect the spatial coherence only. For zero baseline, the spatial coherence must approach $g^{(2)} = 2$, and the autocorrelation function is normalized accordingly since several effects contribute to measured values being lower. A slight deviation from zero baseline occurs because the telescope dimensions are not negligible and their measured spatial coherence corresponds to the convolution of the telescope pupils with the spatial coherence pattern[8]. The ideal value of $g^{(2)} = 2$ is retrieved only for fully linearly polarized light, while in unpolarized light the signal decreases[1] by a factor $\sqrt{2}$ (these measurements were in unpolarized light). Also, the measured signal is diluted by some fraction of coherent laser light that did not scatter against the suspended microparticles but illuminated the aperture by being refracted in the walls of the glass cuvette, preserving its second-order coherence $g^{(2)} = 1$. To account for such effects, the successive cross correlation functions were normalized relative to the autocorrelation ones. Figure 7 shows dependences of the measured second-order coherence on telescopic baseline.

## 3.8 One hundred baselines: Mimicking CTA, the Cherenkov Telescope Array

A number of more 'complex' sources that were measured included asymmetric stars and binary sources of varying sizes and separations. Now the effective number of telescopes and baselines was significantly increased by making successive series of measurements across different angles in the interferometric $(u,v)$-plane. The telescope array is mounted horizontally and thus covers only horizontal baselines. To effectively cover also oblique and vertical baselines, the artificial star was rotated to successively different position angles to change the (*u*,*v*)-plane coverage. (A much simpler procedure than changing the telescope setup or adding numerous additional telescopes on top.) With $N$ telescopes at non-redundant mutual separations, $N(N-1)/2$ different baselines can be constructed. With 5 telescopes, 10 baselines are available at each angular position and, e.g., 10 different position angles produce 100 different baselines. This now moves into new parameter domains for optical interferometry and begins to approach the capabilities of large Cherenkov arrays. Figure 8 shows the coherence pattern for a binary star measured over sixty different baselines.

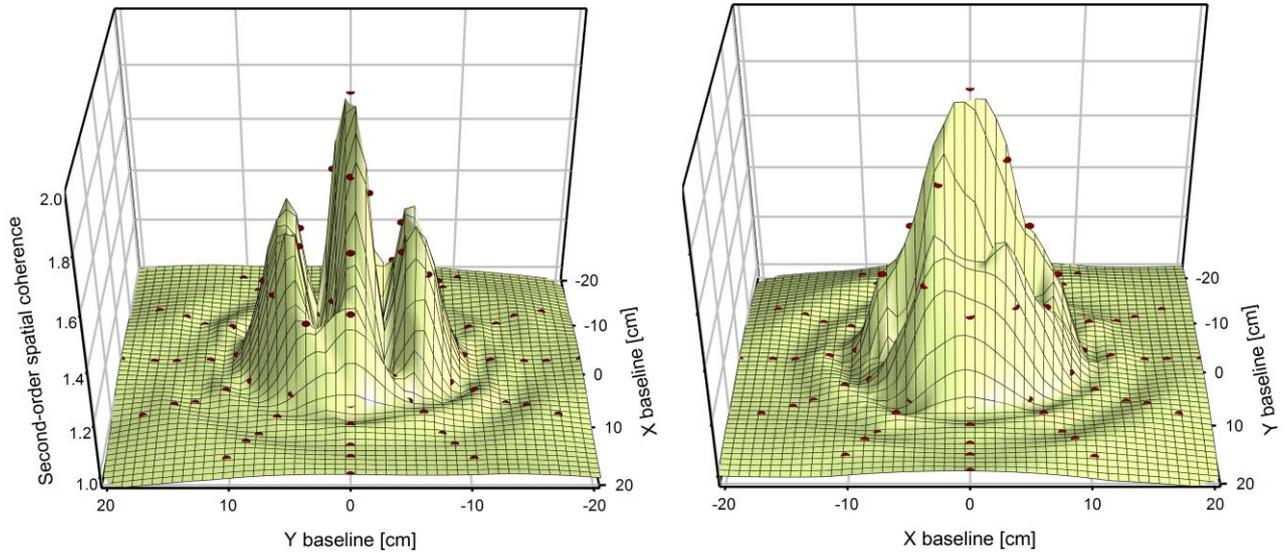

Figure 8. Second-order coherence $g^{(2)}$ measured for an artificial binary star with each component of diameter ~ 1 arcsec. This surface, shown from two orthogonal directions, was produced from intensity correlations measured across 60 different non-redundant baselines, illustrating how a telescope array fills in the interferometric plane. The pattern of central maxima (left) indicates the binary separation while the symmetric rings reveal the size of individual stars, analogous to the patterns in Figure 7. These maps provide the [modulus of the] Fourier transform of the image, analogous to the first-order diffraction patterns produced in coherent light (Figure 2).

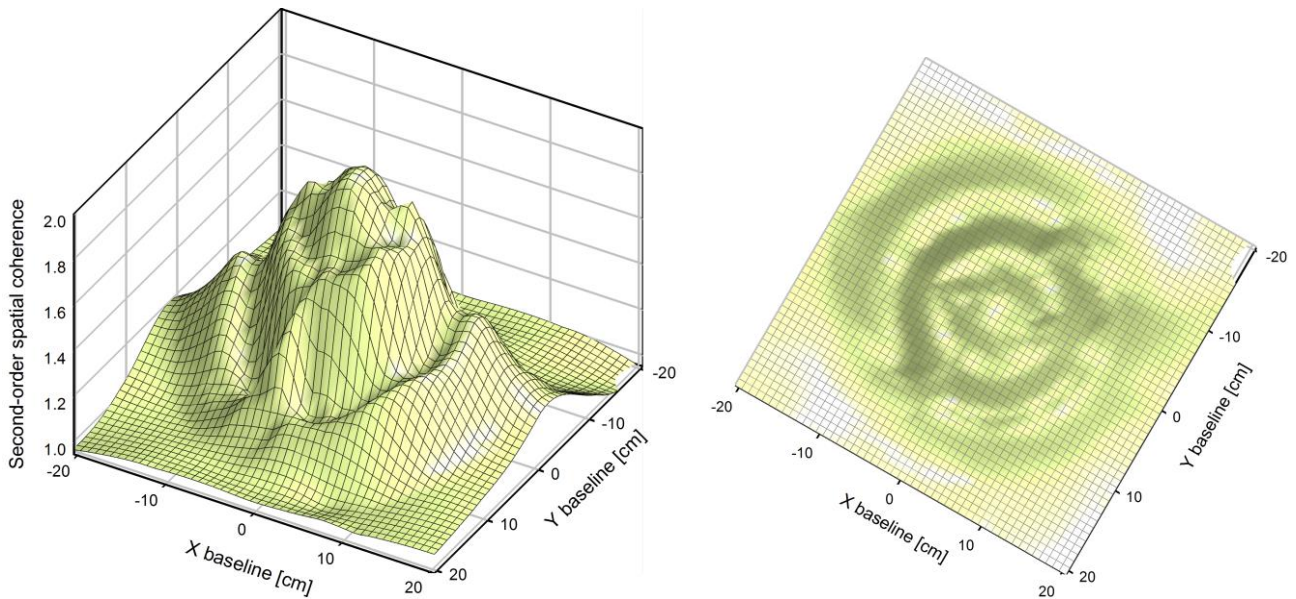

Figure 9. Intensity interferometry measurements with 100 different telescopic baselines. The data largely cover the interferometric (*u*,*v*)-plane of the second-order coherence $g^{(2)}$ for an artificial star, somewhat irregular and elliptic, with angular extent just below 1 arcsecond. At right, the projection of the 3-D mesh is oriented straight down, showing [the modulus of] the source's Fourier transform ('diffraction pattern'). The Cherenkov Telescope Array will provide 1000 baselines, or more.

For Figure 9, the number of non-redundant baselines was increased to 100. These measurements are of a single asymmetric star of a somewhat elliptic and irregular shape, with angular extent just below 1 arcsecond (aperture diameter ~ 80 μm). Assuming 50 telescopes will be available for interferometry with the CTA, the number of baselines, $N(N-1)/2$, would already exceed 1000 (even if some might be redundant due to repetitive telescope locations). As verified in numerical simulations, the ensuing nearly complete $(u,v)$-plane coverage enables full two-dimensional image reconstruction[6-7]. One additional requirement during astronomical observing (although not needed here) will be to electronically and continuously track the changing projected baselines between pairs of telescopes as the source moves overhead across the sky during an observing night, assigning measurements to the instantaneous location in the $(u,v)$-plane.

## 4. TOWARDS KILOMETER-SCALE OPTICAL INTERFEROMETRY

### 4.1 Conclusions from present experiments

This laboratory setup of a multi-telescope intensity interferometer has demonstrated the operation of an intensity interferometer of the Hanbury Brown-Twiss type, involving a hundred baselines and densely covering a large area of the Fourier space in the interferometric $(u,v)$-plane. The photon count rates achieved and handled by the correlators in real time (>1 MHz) are comparable to what can be expected in actual stellar observations, with some spectral feature selected through a narrow bandpass filter. Our currently available correlators can handle signals from up to 20 telescopes simultaneously and, together with experiments carried out with actual Cherenkov telescopes[21-22], this indicates that there seems to be no hindrance for full-scale operations.

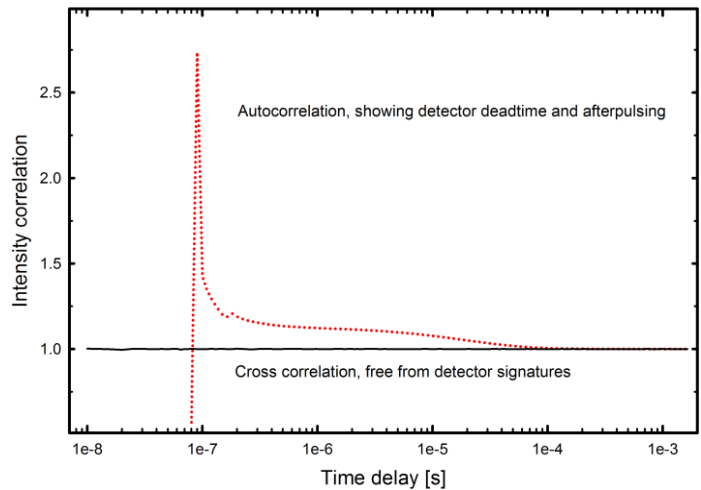

Figure 10. Signatures of detector deadtime and afterpulsing. The autocorrelation of measured intensity in this particular SPAD detector is zero below ~ 80 ns which is its deadtime, within which only one photon can be recorded (the effect also seen as a small bump at twice that value). For the shortest delay times, the autocorrelation rises due to afterpulsing. Although its probability after any one photon detection is small, it stands out strongly in the statistics of averaged signals. For delays beyond ~ 1 μs, the autocorrelation seen here is physical. In cross correlation between the signals from two separate detectors, these instrumental effects vanish (except as a higher-order effect) since afterpulsing is rare and not correlated between different detectors. The physical signal in this particular cross correlation is negligible since those measurements were made with the detectors far apart.

### 4.2 Pending laboratory issues

The current experiments were carried out with thermal light, produced by laser scattering. Although the photon statistics are believed to closely match those of thermal white-light, the coherence times are very much longer than the nanosecond scales relevant for actual astronomical observing. Since coherence times for broader-band light are very much shorter than realistic electronic timescales, most of the cross correlation signal will then be confined to the shortest temporal delays of timescales on a few nanoseconds. Here, various detector issues begin to appear, such as deadtimes and afterpulsing, whose understanding, calibration and mitigation may be important (Figure 10). For example, most probably one will not be able to use the autocorrelation from a single detector to obtain the zero-baseline cross correlation but rather require two detectors side by side (following a beamsplitter) in order to avoid single-detector signatures from dominating the signal. With the aim to realize laboratory experiments on also such short timescales, some tests with high-temperature white-light high-pressure arc lamps have already been made, indicating that it might be feasible to achieve meaningful signal-to-noise ratios although some effort will be required.

For the observability of fainter sources, the signal-to-noise ratio can be improved by simultaneous measurements in multiple spectral bands, exploiting the property that S/N is independent of optical bandpass. This will require some dispersive optics near the focal plane, followed by a multi-element detector, perhaps a SPAD array. Also polarization properties could be optimized: Since the full $g^{(2)}$ signal appears in linearly polarized light, the S/N optimization might be explored by introducing a polarizing beamsplitter and detecting and correlating each polarization separately.

### 4.3 Outlook for field experiments

Cherenkov telescopes normally produce quite large cm-size stellar images in their focal planes, much larger than the single-pixel SPAD detectors used here. Although our detectors are similar in principle to those large-area solid-state photomultipliers now demonstrated in Cherenkov telescopes[26-28], those larger-area ones are built up by a large number of individual light-sensitive areas on the same silicon chip and (as we have already examined in other laboratory experiments) possess somewhat different deadtime, dark-count, and afterpulsing characteristics. An adequate understanding of such detector properties (or of photomultipliers, in case such would be used) will be required to reach the photometric precisions required for disentangling physical values of $g^{(2)}$ from stochastic or systematic noise.

When observing celestial sources during an observing night with stationary telescopes, projected baselines between pairs of telescopes will gradually change. Of course, this enables a richer $(u,v)$-plane coverage and is the principle of aperture synthesis imaging using Earth rotation[10]. However, for real-time correlation, this requires an electronic unit that implements a variable time delay onto the stream of photon pulses, compensating for the relative timings of the wavefront at the different telescopes, as the source moves across the sky[5]. In case the correlation is computed off-line, following a recording of the signal during observation, the variable delay would instead have to be applied in software.

Besides the cross correlation required to retrieve the spatial coherence $g^{(2)}$, there is a potential to explore also higher-order correlations in light. The current intensity interferometer can be seen as one special case of more general spatio-temporal relations in light, measuring the cross correlation between intensity fluctuations at two spatial locations, at one instant in time. However, using telescope arrays, one can construct, e.g., third-order intensity correlations, $g^{(3)}$ for systems of three telescopes: $|<I(\mathbf{r}_1,t_1) \bullet I(\mathbf{r}_2,t_2) \bullet I(\mathbf{r}_3,t_3)>|$, where the temporal coordinates $t$ do not even have to be equal. In principle, such and other higher-order correlations in light carry additional information about the source[45-51]. For instance, from correlations among also all possible triplets and quadruplets of telescopes, a more robust full reconstruction of the source image should be possible. Here, we are entering largely unexplored territory for observational astronomy, but one which could well be enabled by the availability of extended arrays with large and numerous telescopes.

## 5. MICROARCSECOND ASTROPHYSICS

### 5.1 Astronomical sources observable with 'small' Cherenkov telescope arrays

The probably 'easiest' targets for intensity interferometry observations are relatively bright and hot, single or binary stars of spectral types O and B or Wolf-Rayet stars with their various circumstellar emission-line structures. Their stellar disk diameters are typically ~ 0.2–0.5 mas and thus lie (somewhat) beyond what can be resolved with existing amplitude interferometers. Also rapidly rotating stars, with oblate shapes deformed by rotation, circumstellar disks, winds from hot stars, blue supergiants and extreme objects such as η Carinae, interacting binaries, the hotter parts of [super]nova explosions, pulsating Cepheids or other hotter variables are clear candidates[4-5].

Given that the S/N is independent of optical bandwidth, one may as well observe in the light of some specific spectral feature, as in white light. Assuming suitable wavelength filters, one could try to map the non-radial pulsations across the surfaces of stars such as Cepheids. The amplitudes in temperature and white light probably are modest but the associated velocity fluctuations might be observable. If the telescope optics can be sufficiently collimated to permit the use of very narrow-band spectral filters centered on strong absorption lines, the local stellar surface will appear at their particular residual intensity when at rest relative to the observer, but will reach continuum intensity when local velocities have Doppler-shifted the absorption line outside the filter passband. Once such spatially resolved observations of stellar non-radial oscillations are realized, they will certainly provide significant input to models of stellar atmospheres and interiors.

## 5.2 Targets for the full Cherenkov Telescope Array.

The full CTA will have numerous telescopes, distributed over a few square km, with an edge-to-edge distance of two or three km[2-3]. If fully equipped for intensity interferometry at the shortest optical wavelengths, the spatial resolution will approach ~ 30 µas. Such resolutions have hitherto been reached only in the radio, and it is awkward to speculate on what features could appear in the optical. However, to appreciate the meaning of such resolutions, Figure 11 shows an 'understandable' type of object: a hypothetical exoplanet in transit across the star Sirius ($T_{eff}$ = 9,940 K). Its size and oblateness was taken equal to that of Jupiter, but fitted with a Saturn-like ring and four 'Galilean' moons, each the size of Earth. While spatially resolving the disk of an exoplanet in its reflected light may remain unrealistic, the imaging of its dark silhouette on a stellar disk – while certainly very challenging – could perhaps be not quite impossible[52].

Intensity interferometry actually possesses some advantages for such possible observations. The lack of sensitivity to the phases of the Fourier components of the image could be an advantage since one would all the time measure 'only' the amplitude of the Fourier transform of the exoplanet image, irrespective of where on the stellar disk it happened to be. Unless close to the stellar limb, the star only serves as a bright background, its comparatively 'huge' diameter of 6 mas not contributing any sensible spatial power at any relevant telescopic baselines.

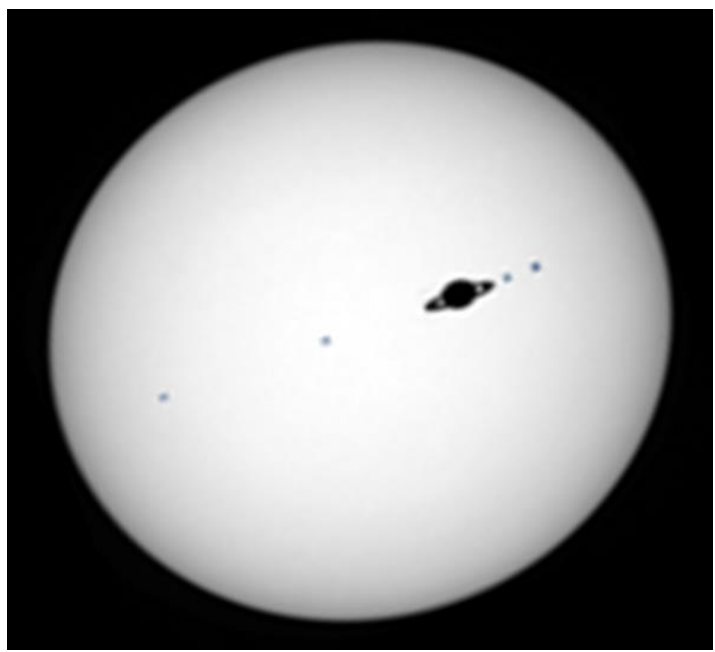

Figure 11. A vision of microarcsecond optical imaging: Expected resolution for an assumed transit of a hypothetical exoplanet across the disk of Sirius, using the Cherenkov Telescope Array as an intensity interferometer. Stellar diameter = 1.7 solar, Distance = 2.6 pc, Angular diameter = 6 mas; assumed planet of Jupiter size and oblateness; Saturn-type rings; four Earth-size moons; equatorial diameter = 350 µas. With the CTA array spanning 2 km, a 50 µas resolution provides more than 100 pixels across the stellar diameter.

## ACKNOWLEDGMENTS

This work is supported by the Swedish Research Council and The Royal Physiographic Society in Lund. The development of concepts for intensity interferometry with Cherenkov telescope arrays has involved interactions with several colleagues elsewhere, in particular at the University of Utah in Salt Lake City (David Kieda, Stephan LeBohec, and Paul D. Nuñez) and at the University of Padova (Cesare Barbieri and Giampiero Naletto). Early experiments towards laboratory intensity interferometry at Lund Observatory involved also Toktam Calvén Aghajani, Hannes Jensen, Ricky Nilsson and Helena Uthas while laboratory studies of SPAD detectors were made by also Daniel Faria and Johan Ingjald. The artificial 'stars' were prepared by the late research engineer Nels Hansson. We also thank the CTA Speakers and Publications Office for constructive refereeing.